\documentclass[12pt]{iopart}

\usepackage{iopams}
\usepackage{graphicx}	
\usepackage{harvard}
\usepackage{xcolor}
\usepackage{hyperref}

\makeatletter

\let\jnl@style=\rmfamily 
\def\ref@jnl#1{{\jnl@style#1}}%
\newcommand\aj{\ref@jnl{AJ}}
\newcommand\araa{\ref@jnl{ARA\&A}}
\newcommand\apj{\ref@jnl{ApJ}}
\newcommand\apjl{\ref@jnl{\@eapj@ApJLetters}}
\newcommand\apjs{\ref@jnl{ApJS}}
\newcommand\ao{\ref@jnl{Appl.~Opt.}}
\newcommand\apss{\ref@jnl{Ap\&SS}}
\newcommand\aap{\ref@jnl{A\&A}}
\newcommand\aapr{\ref@jnl{A\&A~Rev.}}
\newcommand\aaps{\ref@jnl{A\&AS}}
\newcommand\azh{\ref@jnl{AZh}}
\newcommand\baas{\ref@jnl{BAAS}}
\newcommand\icarus{\ref@jnl{Icarus}}
\newcommand\jrasc{\ref@jnl{JRASC}}
\newcommand\memras{\ref@jnl{MmRAS}}
\newcommand\mnras{\ref@jnl{MNRAS}}
\newcommand\pra{\ref@jnl{Phys.~Rev.~A}}
\newcommand\prb{\ref@jnl{Phys.~Rev.~B}}
\newcommand\prc{\ref@jnl{Phys.~Rev.~C}}
\newcommand\prd{\ref@jnl{Phys.~Rev.~D}}
\newcommand\pre{\ref@jnl{Phys.~Rev.~E}}
\newcommand\prl{\ref@jnl{Phys.~Rev.~Lett.}}
\newcommand\pasp{\ref@jnl{PASP}}
\newcommand\pasj{\ref@jnl{PASJ}}
\newcommand\qjras{\ref@jnl{QJRAS}}
\newcommand\skytel{\ref@jnl{S\&T}}
\newcommand\solphys{\ref@jnl{Sol.~Phys.}}
\newcommand\sovast{\ref@jnl{Soviet~Ast.}}
\newcommand\ssr{\ref@jnl{Space~Sci.~Rev.}}
\newcommand\zap{\ref@jnl{ZAp}}
\newcommand\nat{\ref@jnl{Nature}}
\newcommand\iaucirc{\ref@jnl{IAU~Circ.}}
\newcommand\aplett{\ref@jnl{Astrophys.~Lett.}}
\newcommand\apspr{\ref@jnl{Astrophys.~Space~Phys.~Res.}}
\newcommand\bain{\ref@jnl{Bull.~Astron.~Inst.~Netherlands}}
\newcommand\fcp{\ref@jnl{Fund.~Cosmic~Phys.}}
\newcommand\gca{\ref@jnl{Geochim.~Cosmochim.~Acta}}
\newcommand\grl{\ref@jnl{Geophys.~Res.~Lett.}}
\newcommand\jcp{\ref@jnl{J.~Chem.~Phys.}}
\newcommand\jgr{\ref@jnl{J.~Geophys.~Res.}}
\newcommand\jqsrt{\ref@jnl{J.~Quant.~Spec.~Radiat.~Transf.}}
\newcommand\memsai{\ref@jnl{Mem.~Soc.~Astron.~Italiana}}
\newcommand\nphysa{\ref@jnl{Nucl.~Phys.~A}}
\newcommand\physrep{\ref@jnl{Phys.~Rep.}}
\newcommand\physscr{\ref@jnl{Phys.~Scr}}
\newcommand\planss{\ref@jnl{Planet.~Space~Sci.}}
\newcommand\procspie{\ref@jnl{Proc.~SPIE}}

\newcommand\actaa{\ref@jnl{Acta Astron.}}
\newcommand\caa{\ref@jnl{Chinese Astron. Astrophys.}}
\newcommand\cjaa{\ref@jnl{Chinese J. Astron. Astrophys.}}
\newcommand\jcap{\ref@jnl{J. Cosmology Astropart. Phys.}}
\newcommand\na{\ref@jnl{New A}}
\newcommand\nar{\ref@jnl{New A Rev.}}
\newcommand\pasa{\ref@jnl{PASA}}
\newcommand\rmxaa{\ref@jnl{Rev. Mexicana Astron. Astrofis.}}

\makeatother

\begin{document}

\title{Extracting the Optical Depth to Reionization $\tau$  from 21\,cm Data Using Machine Learning Techniques}



\author{
Tashalee S. Billings,$^{1}$\thanks{\href{mailto:tashalee@sas.upenn.edu}{tashalee@sas.upenn.edu}}
Paul La Plante,$^{2,3}$
James E. Aguirre$^{1}$}

\address{$^{1}$Center for Particle Cosmology, Department of Physics \& Astronomy, University of Pennsylvania, Philadelphia PA 19104}
\address{$^{2}$Department of Astronomy, University of California, Berkeley, CA 94720}
\address{$^{3}$Berkeley Center for Cosmological Physics, University of California, Berkeley, CA 94720}


\submitto{PASP}

\begin{abstract}
Upcoming measurements of the high-redshift 21\,cm signal from the Epoch of Reionization (EoR) are a promising probe of the astrophysics of the first galaxies and of cosmological parameters.  In particular, the optical depth $\tau$ to the last scattering surface of the cosmic microwave background (CMB) should be tightly constrained by direct measurements of the neutral hydrogen state at high redshift. A robust measurement of $\tau$ from 21\,cm data would help eliminate it as a nuisance parameter from CMB estimates of cosmological parameters. Previous proposals for extracting $\tau$ from future 21\,cm datasets have typically used the 21\,cm power spectra generated by semi-numerical models to reconstruct the reionization history. We present here a different approach which uses convolution neural networks (CNNs) trained on mock images of the 21\,cm EoR signal to extract $\tau$.  We construct a CNN that improves upon on previously proposed architectures, and perform an automated hyperparameter optimization. We show that well-trained CNNs are able to accurately predict $\tau$, even when removing Fourier modes that are expected to be corrupted by bright foreground contamination of the 21\,cm signal.  Typical random errors for an optimized network are less than $3.06\%$, with biases factors of several smaller.  While preliminary, this approach could yield constraints on $\tau$ that improve upon sample-variance limited measurements of the low-$\ell$ EE observations of the CMB, making this approach a valuable complement to more traditional methods of inferring $\tau$.
\end{abstract}

\maketitle



\section{INTRODUCTION}
\label{sec:introduction}

Perhaps one of the greatest revelations of the study of the universe is that the universe has changed its fundamental state more than once. Less than 180 million years after the Big Bang, the universe broke its silence and the Cosmic Dawn began. This was a tremendous milestone where the first generation of stars and galaxies formed after the Dark Ages. The radiation produced by these first luminous sources profoundly impacted the structure of the intergalactic medium (IGM), and eventually led to the Epoch of Reionization (EoR). Prior to being reionized, the neutral hydrogen gas in the IGM emitted radiation at a wavelength of $\lambda = 21$ cm due to the hyperfine transition of the ground state of the atom. Today, we observe a structured, complicated and diverse universe of stars and galaxies, but very little neutral gas. Knowledge of how the universe transitioned from its neutral state at recombination to its ionized and structured state today is poorly understood. There are several experimental efforts currently underway which will shed light on this portion of the universe's history for the first time.

Upcoming observations of the EoR are expected to be primarily sensitive to astrophysical parameters related to properties of the first stars and galaxies, rather than cosmological parameters such as those inferred from measurements of the cosmic microwave background (CMB). One exception to this is $\tau$, the optical depth to the CMB. Radio interferometry telescopes such as the Hydrogen Epoch of Reionization Array (HERA\footnote{\url{www.reionization.org}}), the Low Frequency Array (LOFAR\footnote{\url{www.lofar.org}}), and the Square Kilometre Array (SKA\footnote{\url{www.skatelescope.org}}) aim to map the thermal distributions and ionization state of neutral hydrogen in the IGM throughout Cosmic Dawn, and will be the only direct probes of the formation of the first generations of stars, galaxies, and stellar-mass black holes.  Full tomographic 3D images generated by these instruments can be useful to learn information about these sources that precipitated reionization.

More imminently, statistical measurements using the 21\,cm power spectrum can provide insight about the EoR. Thus far, measurements of the 21\,cm power spectrum upper limits have been established at different Fourier wavenumbers and redshift values \cite{paciga_etal2013,beardsley_etal2016,patil_etal2017,kolopanis_etal2019}. However, direct extraction of astrophysical and cosmological parameters from the the power spectrum or from images is challenging due to large levels of contamination from bright foreground emission which are typically several orders of magnitude larger than the target signal. This limitation makes simple imaging of the sky using traditional techniques impossible. The main difficulty in detecting the faint signal from the EoR is to separate it from various types of foreground emissions such as galactic synchrotron radiation and extragalactic point sources \cite{dimatteo_etal2004,jelic_etal2008}. Another common approach proposed for extracting information about the EoR from observations is to compute the power spectrum using Fourier modes that are uncontaminated by these foregrounds.
The downside to any power-spectrum based approach that is is insensitive to any non-Gaussian information and therefore does not leverage all of the information 
present in the images. The power spectrum does not capture the full information present in the field because the 21\,cm field is highly non-Gaussian during the EoR \cite{Majumdar:2017tdm,Shimabukuro:2017jdh}.

Using measurements of the 21\,cm signal, it may be possible to infer key properties of the EoR, such as the timing and duration of reionization. Although these properties are interesting in their own right, they also provide important information about the cosmological parameter $\tau$, which measures the optical depth to the CMB. To date $\tau$ has been measured by the Planck collaboration, which has provided important constraints on its value. However, the value of $\tau$ still has some of the largest relative uncertainty of the cosmological parameters, which impacts the uncertainty of other parameters such as the density of dark matter $\Omega_{c}$ and clustering of matter $\sigma_{8}$. \cite{liu16} proposed using measurements of the 21\,cm power spectrum as a way to provide tighter constraints than is currently feasible from CMB measurements alone, which can lead to improved uncertainties of other parameters. The authors jointly constrained $\tau$ and other cosmological parameters using semi-numeric simulations of reionization, leading to a fractional uncertainty several times better than current CMB-based constraints. Thus, data analysis techniques that provide constraints on $\tau$ are promising ways forward for measuring cosmological parameters more accurately.

An alternative approach to computing power spectra is to use supervised machine learning techniques on simulated image cubes of the EoR by using two-dimensional convolution neural networks to perform regression on astrophysical and cosmological parameter values, and ultimately predict on new images not previously seen by the network and predict the desired reionization values. This image processing approach allows for the extraction of non-Gaussian information present in the maps. In this paper we discuss our approach to extracting $\tau$ using convolution neural networks (CNNs).

Machine learning techniques have been exploited in a variety of fields to explore different scientific questions. For example, the authors of \cite{hortua_etal2019} use approximate Bayesian Neural Networks (BNNs) to predict the posterior distribution of the cosmological parameters directly from the CMB temperature and polarization maps. In the context of 21\,cm data, \cite{gillet_etal2018} used CNNs to extract semi-analytic model parameters related to astrophysics from \textsc{21cmfast} simulations. \cite{laplante_ntampaka2019} applied CNNs to simulated images of the EoR, and were able to successfully infer the duration of reionization to a high degree of accuracy. There have also been several recent studies where cosmological or astrophysical parameters are inferred by applying machine learning techniques to simulated 21\,cm data \cite{zamudio-fernandez_etal2019},\cite{makinen_etal2020,kwon_etal2020},\cite{villanueva-domingo_etal2020,wadekar_etal2020}. In this paper, we build upon the approach of \cite{laplante_ntampaka2019}, and vary the reionization history to include changes in the midpoint and duration of reionization. We also predict directly on $\tau$, rather than inferring the reionization meta-parameters. This approach allows us to compare more directly with the uncertainty on $\tau$ related to other methods.


This paper is organized in the following manner: in Sec.~\ref{sec:data}, we describe the reionization model used and the method to generate the input image cube. In Sec.~\ref{sec:cnn}, we describe the machine learning approaches used, as well as present our results. In Sec.~\ref{sec:discussion}, we further discuss interpretations of our results and compare with other methods of inferring $\tau$. In Sec.~\ref{sec:conclusion}, we conclude and discuss future research. Throughout this work, we assume a $\Lambda$CDM cosmology with parameters consistent with the Planck 2018 results \cite{planck2018}.

\section{Data Design}
\label{sec:data}

The simulated 21\,cm data used in this paper was generated using the semi-numeric technique first developed in \cite{battaglia_etal2013a}.  
This model considers the redshift at which different region in the universe become highly ionized, such that the ionization fraction $x_i \sim 1$. This leads to defining a local ``redshift of reionization'' field $z_\mathrm{re}(\textbf{x})$, with fractional fluctuations $\delta_z(\textbf{x})$:
\begin{equation}
    \delta_z(\textbf{x}) = \frac{\left[z_\mathrm{re}(\textbf{x}) + 1\right] - \left[\bar{z} + 1\right]}{\bar{z} + 1},
\label{eq:redshift}
\end{equation}
where $\bar{z}$ is the mean redshift of reionization, chosen as an input to the model. 
The reionization field $\delta_z(\textbf{x})$ is assumed to be a biased tracer of dark matter on large scales ($\geq 1$ $h^{-1}$Mpc) with 
bias parameter $b_{zm}(k)$:
\begin{equation}
  b^2_{zm}(k) \equiv \frac{\left\langle \delta_z^* \delta_z \right\rangle}{\left\langle \delta_m^* \delta_m \right\rangle} = \frac{P_{zz}(k)}{P_{mm}(k)}.
\end{equation}
This bias parameter is a three-parameter function of Fourier wavenumber $k$ 
and the result of relating the dark matter density and redshift fields:
\begin{equation}
    b_{zm}(k) = \frac{b_0}{\left(1 + \frac{k}{k_0}\right)^{\alpha}},
\label{eq:biasfunction}
\end{equation}
where $b_0$ is the bias amplitude, $k_0$ is the scale threshold, and $\alpha$ is an asymptotic exponent. We use a value of $b_0 = 1/\delta_c = 0.593$. Given this parameterization, we are able to vary the reionization history by changing the parameter $\bar{z}$ to modify the midpoint, and the parameters $k_0$ and $\alpha$ to adjust the duration.

The dark matter density field is generated at the mean redshift $\bar{z}$, then it is Fourier transformed into $k$-space. The bias function in Equation~(\ref{eq:biasfunction}) is then used to generate $\delta_z(\textbf{k})$ by simple mode-wise multiplication. An inverse Fourier transform is applied to this $k$-space field to arrive at $\delta_z(\textbf{x})$. Then it is finally inverted using Equation~(\ref{eq:redshift}) to get the field $z_\mathrm{re}(\textbf{x})$, the reionization history for some volume. We can use the redshift of reionization field to calculate the local ionization field for some redshift $z$. Finally, we combine the local ionization field with the matter density field to compute the 21\,cm signal:

\begin{eqnarray}
\eqalign{
    \delta T_b &= 26(1 + \delta_m) x_{\mathrm{H\sc{I}}} \left(\frac{T_S - T_{\gamma}}{T_S}\right) \left(\frac{\Omega_bh^2}{0.022}\right) \\
    &\qquad \times \left[\left(\frac{0.143}{\Omega_mh^2}\right)\left(\frac{1 + z}{10}\right)\right]^{\frac{1}{2}}\, \mathrm{mK}}
\label{eq:21cmtemp}
\end{eqnarray}

where $x_{\mathrm{H_I}} = 1 - x_i$ is the neutral fraction field for a given point in
the volume, $T_S$ is the spin temperature of the gas, and $T_{\gamma}$ is the temperature of the CMB at some redshift. Using this semi-analytic  model of reionization, we can generate mock images of the 21\,cm brightness temperature $\delta T_b$ at different redshift values in an efficient way by adjusting the parameters $\bar{z}$, $k_0$, and $\alpha$. 

For this study, we generated 1000 realizations of the 21\,cm field from a dark-matter density field generated from a single $N$-body simulation which tracked 2048$^3$ particles in a cubic volume of 2~$h^{-1}$Gpc on a side using a P$^3$M algorithm described in \cite{trac_etal2015}.  To avoid presenting the same density structures in different 21\,cm realizations (and thus potentially biasing the results), the line-of-sight direction and zero-indexed pixels were randomly chosen. Once the starting indices for each axis were chosen, the box was permuted using periodic boundary conditions. This approach helps mitigate repetition of the underlying density field for the purposes of generating snapshots. We then randomly sample the parameters $\bar{z}$, $\alpha$, and $k_0$ from a uniform range of values to generate a unique reionization field $z_\mathrm{re}(\textbf{x})$.\footnote{Specifically, we chose values of $7 \leq \bar{z} \leq 9$, $0.3 \leq \alpha \leq 0.9$, and $0.07 \leq k_0/(\mathrm{Mpc^{-1}}h) \leq 0.35$.}
In order to obtain the density field at $z = \bar{z}$, the two neighboring matter density fields are
interpolated in scale factor $a$ for every point in the volume.
This allows for the construction of an approximate density field for any desired redshift without having to run a new simulation.  The simulation of that particular realization then proceeds as outlined above. 

Note that this method does not take $\tau$ directly as an input, but given the field $z_\mathrm{re}(\textbf{x})$, the value of $\tau$ can be computed from the simulated volume.
In general, the reionization histories produced by our parameter choices feature a late end of reionization, and tend to have a relatively broad duration of reionization. The corresponding values of $\tau$ range from $0.045 \leq \tau \leq 0.068$. This range covers the values of $\tau$ reported by the Planck 2015 and Planck 2018 cosmological parameters. Afterwards, 30 redshift values at constant intervals in co-moving distance between $6 \leq z \leq 12$ are chosen. Two-dimensional slices are generated at each redshift value, which serve as the input data for the CNN architecture. By treating these 30 input images as ``color channels'' in the input data, we are able to make full use of the tomographic data potentially available from observations. 


To understand the performance of the CNN in the presence of foreground contamination, we generate two versions of the same input data: one using just the data from the simulation, and another where the modes expected to be contaminated by foreground emission have been removed from the data. The power of foreground emission can be written as a function of the Fourier mode along the line of sight $k_\parallel$ and in the plane of the sky $k_\perp$. The slope $m$ relating the two is a function of redshift, but is largely independent of instrument specifics. The boundary between the foreground contaminated and foreground free region is given by
\cite{thyagarajan_etal2015}:

\begin{equation}
m(z) \equiv \frac{k_\parallel}{k_\perp} = \frac{\lambda(z) D_c(z) f_{21} H(z)}{c^2 (1+z)^2},
\label{eq:slope}
\end{equation}

where $\lambda(z) = \lambda_0(1+z)$ is the wavelength of the 21\,cm radiation at the redshift of interest, $D_c$ is the co-moving distance to redshift $z$, $f_{21}$ is the rest-frame frequency of the 21\,cm signal, and $H$ is the Hubble parameter. For the redshifts of interest, $m \sim 3$.  We 
approximate the effects of ignoring contaminated foreground modes
by setting all modes below the slope $m$ in Equation~(\ref{eq:slope}) to 0. For this ``Cut'' input data where the foreground-contaminated modes were removed, we extracted a slab of 50 pixels along the line of sight (so the full slab had dimensions of 2048 $\times$ 2048 $\times$ 50) and applied the wedge cut individually to each of the 30 input slabs of data. Afterwards, we selected the central slice from the input data to serve as a representative sampling of the slab.

As a point of comparison, we also generated ``Full'' input data which did not remove any $k$-modes, and the 30 input slices were merely averages over this same 50-pixel slab. In both cases, we also downsampled from the native 2048 $\times$ 2048 pixel resolution within a slice to 512 $\times$ 512 pixels, in order to make the data more manageable for the machine learning application. Note that in practice the actual combination of different co-moving regions along the line of sight will be determined by the observing strategy of the instruments, though for the time being we use this approach as an approximation. Also note that this approach does not include other sources of observational uncertainty, such as thermal noise from the instrument or other systematic errors. We defer additional treatment of these issues to future work.

\section{Convolution Neural Network Regression}
\label{sec:cnn}


In this section, we discuss the machine learning techniques used in this analysis. As mentioned above in Sec.~\ref{sec:introduction}, CNNs and other machine learning techniques have been shown to be well-suited to image-based regression and classification problems. (See e.g., \cite{baron2019}, \cite{ntampaka_etal2019}, and \cite{villaescusa-navarro_etal2020}) We present here the methods used for applying CNNs to the problem of inferring the value of $\tau$ from simulated images of reionization. In Sec.~\ref{sec:architecture}, we discuss the specifics of the CNN networks used. In Sec.~\ref{sec:biasvar}, we discuss the bias-variance tradeoff and a particular manifestation of it in this application. In Sec.~\ref{sec:hyperparameters}, we describe how the technique of hyperparameter optimization allowed us to discover models with much improved performance compared to a previous application in \cite{laplante_ntampaka2019}. In Sec.~\ref{sec:kfold}, we outline our use of the $k$-fold validation technique to demonstrate reliable training and prediction accuracy for our models.  Finally, in Sec.~\ref{sec:results}, we present the results of training our CNN on the input data.

\subsection{Network Architecture}
\label{sec:architecture}
\begin{figure*}
    \centering
    \includegraphics[width=1.0\linewidth]{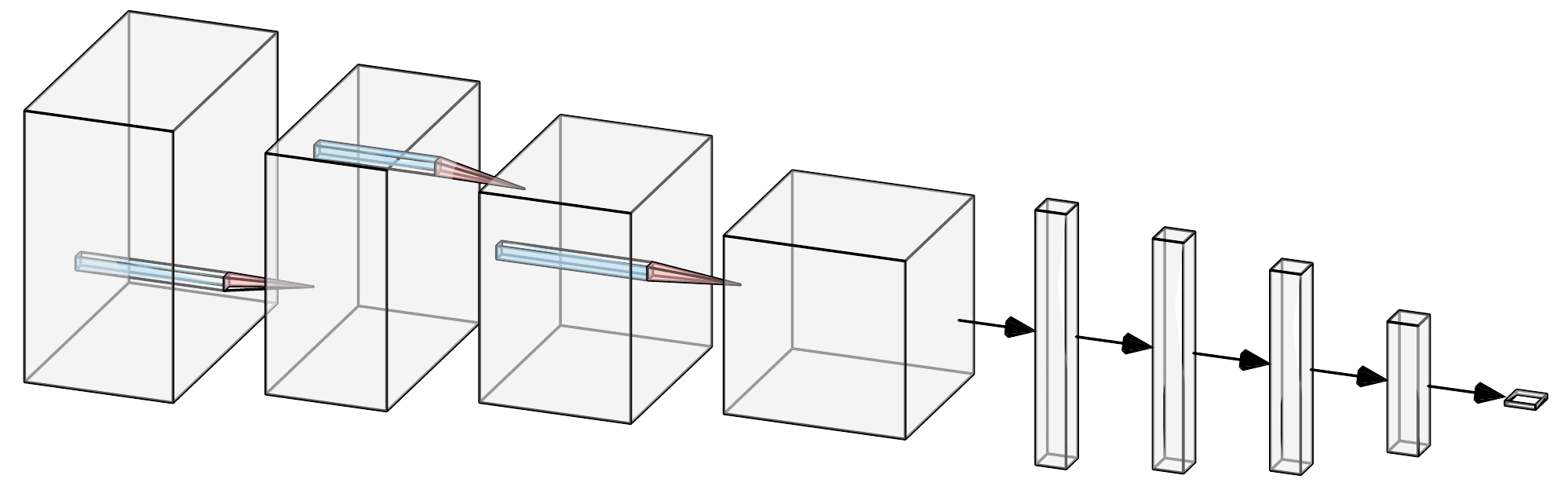}
    \caption{A visualization of a typical convolution neural network (CNN) architecture used in the analysis.  Cubes represent convolutional layers and vertical bars fully connected layers. Note that other layers, such as max pooling or regularization layers, are not explicitly depicted in the figure.
    The specific model shown is the  ``Full Modes'' CNN model (Table \ref{table:ModelSummary}). The image was generated using the \texttt{NN-SVG} tool. 
    The input images are $512 \times 512 \times 30$, and the output is a vector value of length one, corresponding to the optical depth $\tau$.
     }
    \label{fig:CNNArchitecture}
\end{figure*}

Two-dimensional convolutional neural networks (CNNs) are deep learning algorithms that take in some input data, use an algorithm such as backpropagation to adjust weights and biases internal to the network, and then typically solve regression or classification problems. 
Figure~\ref{fig:CNNArchitecture} shows a visualization of
a network used in this work.  The CNN layers are shown as cubes, with skewers through the cube depicting the convolutions, and dense layers are columns. The diagram represents a number of CNN hidden layers, after which the transition is made to dense layers. The final output layer is a single neuron containing the value of $\tau$. 
In general, the CNN architectures used in this work begin with \textbf{two-dimensional convolution} layers
with a stride of one followed by the \textbf{rectified linear unit activation} (ReLU) function. The purpose of the activation function is to become ``active'' and transfer data exiting one neuron to the next. This non-linearity is key to successful operation of the machine learning network. If the neuron is not activated, no information gets through. After that comes a \textbf{batch normalization layer} \cite{ioffe_szegedy2015,santurkar_etal2018} and finally, a two-dimensional \textbf{max pooling layer} with a stride of 2 (e.g., \cite{riesenhuber_poggio1999}). The batch normalization layer works to restrict the activation of each layer to strictly have zero mean and variance of one.
This was once called ``covariate shift'' and if ignored it can be a problem because the behavior of machine learning algorithms can change when the input distribution changes from layer to layer. It is important to limit covariate shift by normalizing the activation neurons of each layer and as a result batch normalization transforms the inputs to be mean zero and variance of value one making them constant.

Pooling layers allow for down sampling important features within an image by summarizing the presence of features into patches of the feature map. This pooling produces a feature image with low resolution. \textbf{Average pooling} and \textbf{max pooling} summarize the average presence of a feature and the most activated presence of a feature respectively. In our networks, max pooling layers are repeated four times. Then, the two-dimensional \textbf{global average pooling} layer is used. After this, the network alternates using dropout layers and dense layers four times before it reaches the output layer, the predicted value of $\tau$ based on the input images. The \textbf{dropout} (20\% dropout) layer is a regularization method that randomly ignores some number of neurons in some layer outputs during the training process \cite{srivastava_etal2014}. This dropout is done according to the Bernoulli distribution.
The \textbf{dense} or \textbf{fully-connected} layer, takes all the input features from different neurons and makes them connected to all the neurons in each layer. 

For defining, training, and predicting using our CNN architectures, we make use of Keras. Keras \cite{chollet2015keras} is a high-level wrapper of TensorFlow \cite{abadi_etal2016}, a numerical library capable of constructing and training machine learning networks. It is important to point out that the weights and biases are considered ``trainable parameters,'' and are updated during the backpropagation process by some optimization algorithm.


\subsection{Bias-Variance Tradeoff}
\label{sec:biasvar}

\begin{figure*}
\centering
\includegraphics[width=0.49\linewidth]{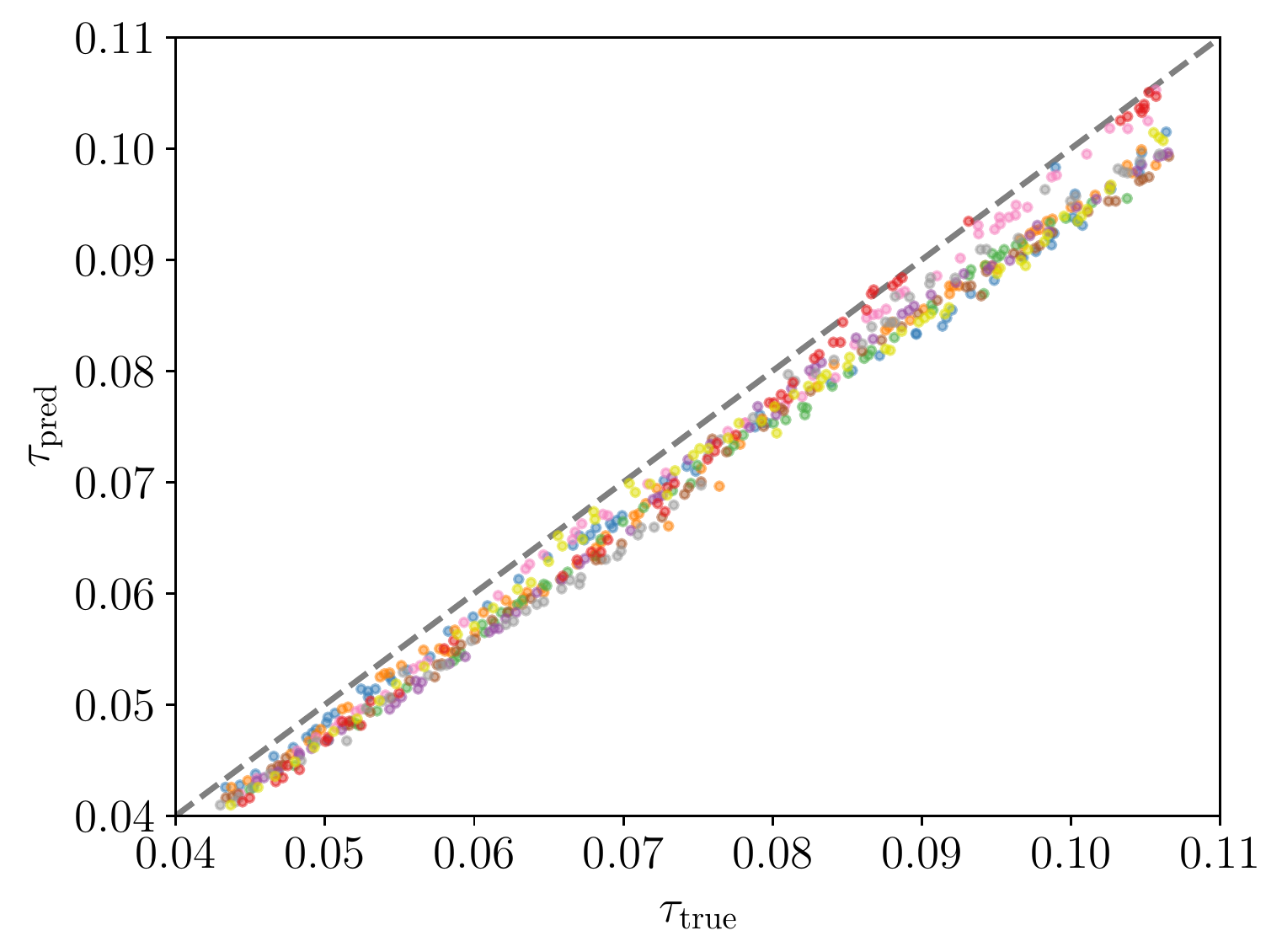}\hfill
\includegraphics[width=0.49\linewidth]{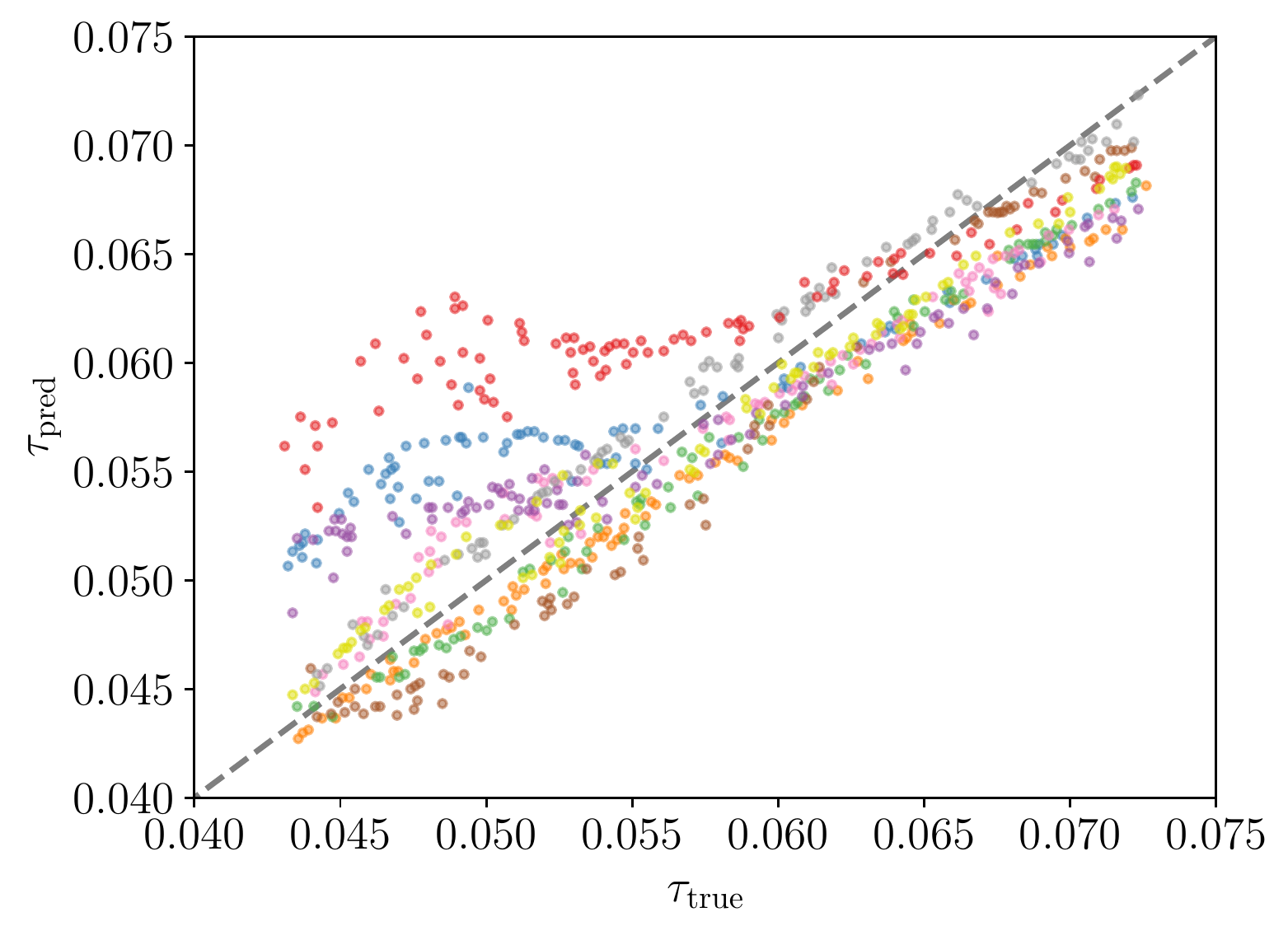}
\caption{A visualization of the bias/variance trade-off for two different networks. These networks were small perturbations of the model used in \protect\cite{laplante_ntampaka2019}.
At left is a model with low variance but significant bias, and at right one with large heteroscedastic variance and high bias. As can be seen, the error in the predicted value is quite large, which suggests that additional complexity is needed to generate accurate prediction. See further discussion in Sec.~\ref{sec:biasvar}.}
\label{fig:bias_var_ex}
\end{figure*}

When designing a machine learning model, one must contend with bias, variance, and noise error components. In general, one can write the total loss as a combination of these three components. This well-known result is known as the ``Bias-Variance Tradeoff''. (For a derivation and examples, see \cite{rajnarayan_wolpert2008} and references therein.) For the sake of brevity, we merely quote the primary result here. We assume we have some data points $(\mathbf{x}_n,y_n)$, where $\textbf{x}_n$ are $n$ different feature vectors and $y_n$ are $n$ different labels. A randomly selected subset of these points are chosen to generate the training set $D = \{(\mathbf{x}_1, y_1), \dots, (\mathbf{x}_n,y_n)\}$. We write the expectation value of the labels as $\bar{y} = E_D[y]$. An algorithm $\mathcal{A}$ trained on these data yields a hypothesis $h_D \equiv \mathcal{A}(D)$, whose expected value is $\bar{h} = E_D[h_D]$. The overall loss function can be expressed as the sum of three distinct components:
\begin{equation}
\eqalign{
    E_{\mathbf{x}, y, D} \left[\left(h_{D}(\mathbf{x}) - y\right)^{2}\right] &= \underbrace{E_{\mathbf{x}}\left[\left(\bar{h}(\mathbf{x}) - \bar{y}(\mathbf{x})\right)^{2}\right]}_\mathrm{Bias^2} \nonumber\\
    &\quad + \underbrace{E_{\mathbf{x}, D}\left[\left(h_{D}(\mathbf{x}) - \bar{h}(\mathbf{x})\right)^{2}\right]}_\mathrm{Variance} \nonumber \\
    &\quad + \underbrace{E_{\mathbf{x}, y}\left[\left(\bar{y}(\mathbf{x}) - y\right)^{2}\right]}_\mathrm{Noise}}
\label{eq:finalgeneralizationerror}
\end{equation}
This partitioning of the overall error function into these three components has interesting implications for the performance of a machine learning network. For instance, a trained network with a comparatively small variance but a consistent bias can yield the same loss value as an unbiased network that has a larger variance. In addition to the bias and variance, the loss function is also sensitive to the noise inherent to the dataset (sometimes called aleatoric uncertainty). Although ideally a trained network has small bias and small variance, during the training process the network can fall into local minima where additional training yields values with high bias and small variance, and further attempts to decrease the loss function merely further reinforce biased predictions.

Figure~\ref{fig:bias_var_ex} shows an example of some initial results after training a CNN model using input data. The two results shown are for networks that were small perturbations of the final network used in \cite{laplante_ntampaka2019}. In the previous work, the authors used a CNN to infer the midpoint and duration of reionization, with the midpoint being held fixed. In this work, the input data varies in both the midpoint and duration. As can be seen in the figure, the network did not perform adequately when predicting the value of $\tau$. The figure shows the predicted value for $\tau$ in a validation set of data versus the true value for a fully trained network. The different colors show the results for different folds. In the left panel, the predictions show relatively small variance, but a high bias. In the right panel, there is high variance and high bias, especially for small values of $\tau$ for certain folds. The variance is also clearly heteroscedastic (i.e., the variance depends on the value of $\tau$). A common method used to evaluate whether the network is overfitting or underfitting is by looking at how the loss behaves as a function of training steps. However, this method is not sensitive to the relative trade-off between bias and variance, which must be examined explicitly. Possible reasons for the worse performance include the input data covering a wider range of possible reionization scenarios, as well as the input data having more 2D slices (30 in this work versus 20 in the previous).

These features when training a neural network can be interpreted as a manifestation of the bias-variance tradeoff. In effect, rather than predicting values that are unbiased and have variance induced by the uncertainty in training and noise, the network predicts values that have high bias with low variance, or ones that have bias values inversely related to the true values (e.g., values that are biased high for small input values, and are biased low for high input values). In general, the poor performance of a neural network (either high bias, high variance, or both) may be fixed by adding complexity to the network \cite{geman_etal1992}. However, determining the best way in which to add the requisite complexity is not a straightforward task. This complexity is governed by so-called hyperparameters, such as the number of hidden layers and their shapes. We now turn to the problem of determining the optimal combination of hyperparameters, and the effect it has on the overall results.

\subsection{Model Hyperparameter Optimization}
\label{sec:hyperparameters}

\begin{table}
\centering
\begin{tabular}{||c c||}
  \hline
  Parameter & Values \\[0.5ex]
  \hline\hline
  Learning Rate & [0.1, 0.01, 0.001, 0.0001] \\
  Loss Function & [RSR, MSE, MAE, MAPE, MSLE]\\
  Number of Convolution Layers & [3, 4, 5]\\ 
  Convolution Filter Size & [125, 256] \\
  Dense Layer & [(200-350, 200, 100, 20)] \\
  \hline
  Batch Size & 32 \\
  Optimizer & \textsc{Adam} \\
  Activation Function & ReLU \\
  Dropout Rate & 20\% \\
  Metrics & [loss, validation loss] \\
  \hline
\end{tabular}
\caption{A summary of the parameters used in the final models. Parameters in the top half of the table were included in the hyperparameter optimization. The different loss functions and ultimate best options are explained further in Section~\ref{sec:hyperparameters}. Parameters in the bottom half of the table were not included in the hyperparameter optimization, and fixed to the values shown.} 
\label{table:OptimizedHyperparameters}
\end{table}


\begin{figure}
    \centering
    \includegraphics[width=1.0\textwidth]{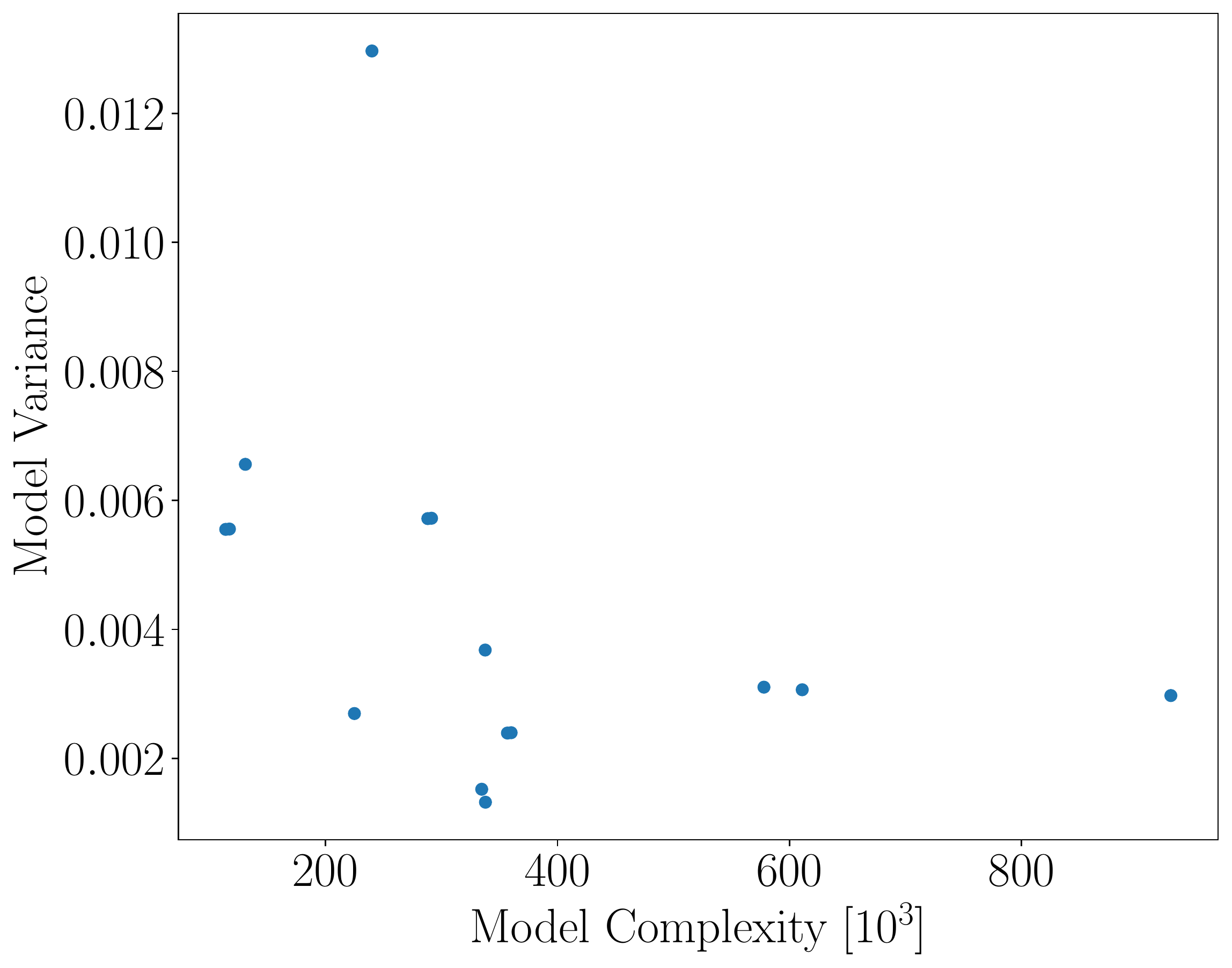}
    \caption{In order to select the best possible model from the parameter tuning, we evaluated 15 different models on 10 different test data and calculated the variance of the error for each of thee 15 models. We essentially picked the best model by prioritising first the model with the smallest variance and then the lowest complexity (number of trainable parameters). This method was applied for models trained on both the ``Full'' data (shown here) as well as the ``Cut'' data.}
    \label{fig:ModelComplexity}
\end{figure}

In the world of machine learning the goal of any supervised machine learning model is to minimize the loss function while still remaining general. To achieve this state, the optimal model parameters and hyperparameters must be selected. Model parameters, sometimes referred to as ``weights and biases,'' are variables whose best values are informed by the data. These variables are updated during the training process. 
By updating these parameters, the optimization algorithm indirectly assigns importance to certain features that minimize the loss. In contrast, model hyperparameters are variables associated with the model that are determined only once, prior to the training process. Put another way, these hyperparameters are not typically modified as part of the training process using the input data. At the same time, choosing appropriate values for these hyperparameters is essential for generating results that are acceptably accurate. When attempting to systematically infer which hyperparameter choices are best for a given application, there are several different strategies available.

Most systematic hyperparameter optimization techniques require first constructing a ``grid'' of hyperparameters. The user specifies the hyperparameters the algorithm has the freedom to vary, along with an array of acceptable values they each can take on. The outer product of all such combinations forms a multi-dimensional grid of choices. We made use of the \texttt{Keras-Tuner}\footnote{\url{https://keras-team.github.io/keras-tuner/}} package to perform the hyperparameter optimization. Specifically, we used the so-called random search method, proposed by \cite{bergstra_bengio2012}. We built and optimized two different networks for the ``Full'' and ``Cut'' datasets. Though initially we tried using a single network for both, we abandoned this approach as data trained on one dataset performed poorly when validated on the other.



Table~\ref{table:OptimizedHyperparameters} displays hyperparameters used as part of the grid search for the hyperparameter optimization, as well as parameters that were fixed. The top half of the table shows the hyperparameters that were allowed to vary. In particular, it includes various choices for the loss function. These  were: relative square residual (RSR), mean square error (MSE), mean absolute error (MAE), mean absolute percentage error (MAPE), and mean squared logarithmic error (MSLE). Mathematically, RSR can be expressed as:
\begin{equation}
    \mathrm{RSR} = \left(\frac{y_\mathrm{true} - y_\mathrm{pred}}{y_\mathrm{true}}\right)^2
\label{eq:RSR}
\end{equation}
Note that in addition to quantities used by the optimizer, such as the learning rate and the loss function, we also varied the architecture itself: we allowed the number of convolution layers to vary, as well as the number of neurons in the dense layers. In addition to these hyperparameter which were varied, the bottom half of Table~\ref{table:OptimizedHyperparameters} shows auxiliary parameters that were fixed as part of the optimization process. In general, these parameters did not have a significant impact on the overall performance of the network, and so we held them fixed while varying other quantities. Additionally, due to the relatively small value of $\tau$, we rescaled the labels for the testing and training datasets by a factor of 1000, which was then removed from predicted values. This led to faster convergence when training the networks, especially for loss functions such as MSE that do not normalize by the ``true'' input values.

\begin{table}
\centering
\begin{tabular}{||c c c||}
    \hline
   La Plante \& Ntampaka & Full Modes & Cut Modes \\ [0.5ex] 
    \hline\hline
    16 3x3 Conv2D filters & 16 3x3 Conv2D filters & 16 3x3 Conv2D filters \\
    BatchNormalization & BatchNormalization & BatchNormalization \\
    2x2 MaxPooling2D & 2x2 MaxPooling2D & 2x2 MaxPooling2D \\
    32 3x3 Conv2D filters & 32 3x3 Conv2D filters & 32 3x3 Conv2D filters \\
    BatchNormalization & BatchNormalization & BatchNormalization \\
    2x2 MaxPooling2D & 2x2 MaxPooling2D & 2x2 MaxPooling2D \\
    64 3x3 Conv2D filters & 64 3x3 Conv2D filters & 64 3x3 Conv2D filters \\
    BatchNormalization & BatchNormalization & BatchNormalization \\
    2x2 MaxPooling2D & 2x2 MaxPooling2D & 2x2 MaxPooling2D \\
    --- & 256 3x3 Conv2D filters & 128 3x3 Conv2D filters \\
    --- & BatchNormalization & BatchNormalization \\
    --- & 2x2 MaxPooling2D & 2x2 MaxPooling2D \\
    --- & --- & 128 3x3 Conv2D filters \\
    --- & --- & BatchNormalization \\
    --- & --- & 2x2 MaxPooling2D \\
    GlobalAvgPooling2D & GlobalAvgPooling2D & GlobalAvgPooling2D \\
    --- & 20\% Dropout & 20\% Dropout \\
    --- & 350 neurons FC & 250 neurons FC \\
    20\% Dropout & 20\% Dropout & 20\% Dropout \\
    200 neurons FC & 200 neurons FC & 200 neurons FC \\
    20\% Dropout & 20\% Dropout & 20\% Dropout \\
    100 neurons FC & 100 neurons FC & 100 neurons FC \\
    20\% Dropout & 20\% Dropout & 20\% Dropout \\
    20 neurons FC & 20 neurons FC & 20 neurons FC \\
    Output neuron &Output neuron & Output neuron \\[1ex] 
    \hline
\end{tabular}
\caption{A summary of the model with the number of parameters expressed using Keras, a high-level python deep learning library that uses a TensorFlow/Theano backend to do lower level calculations. The model on the far left was trained in \protect\cite{laplante_ntampaka2019}, the center model was optimized using the full data without the foreground-contaminated $k$-modes removed, and the model on the right was optimized and trained on data with the foreground $k$-modes filtered out.}
\label{table:ModelSummary}
\end{table}

In order to evaluate the best overall network, we used two main criteria. The first one used was the value of the loss function: given a fully trained network, we examined the average value of the loss function across the validation data. There were several models that performed noticeably worse than the others. These poorly performing networks tended to be less complex, in the sense that they contained fewer trainable parameters. Above a particular number of parameters, many of the networks performed comparably in terms of the average loss value. This led to the second criterion used: the complexity of the network. We quantified the complexity by looking at the number of trainable parameters in the network. Thus, the ``best'' network chosen was the one that had the smallest number of trainable parameters while still performing well in terms of the average loss function. We performed the random search separately for the two different sets of input data, and found that slightly different network architectures yielded the best results. We talk more about these results below in Sec.~\ref{sec:results}.

\begin{figure*}
  \centering
  \includegraphics[width=1.0\textwidth]{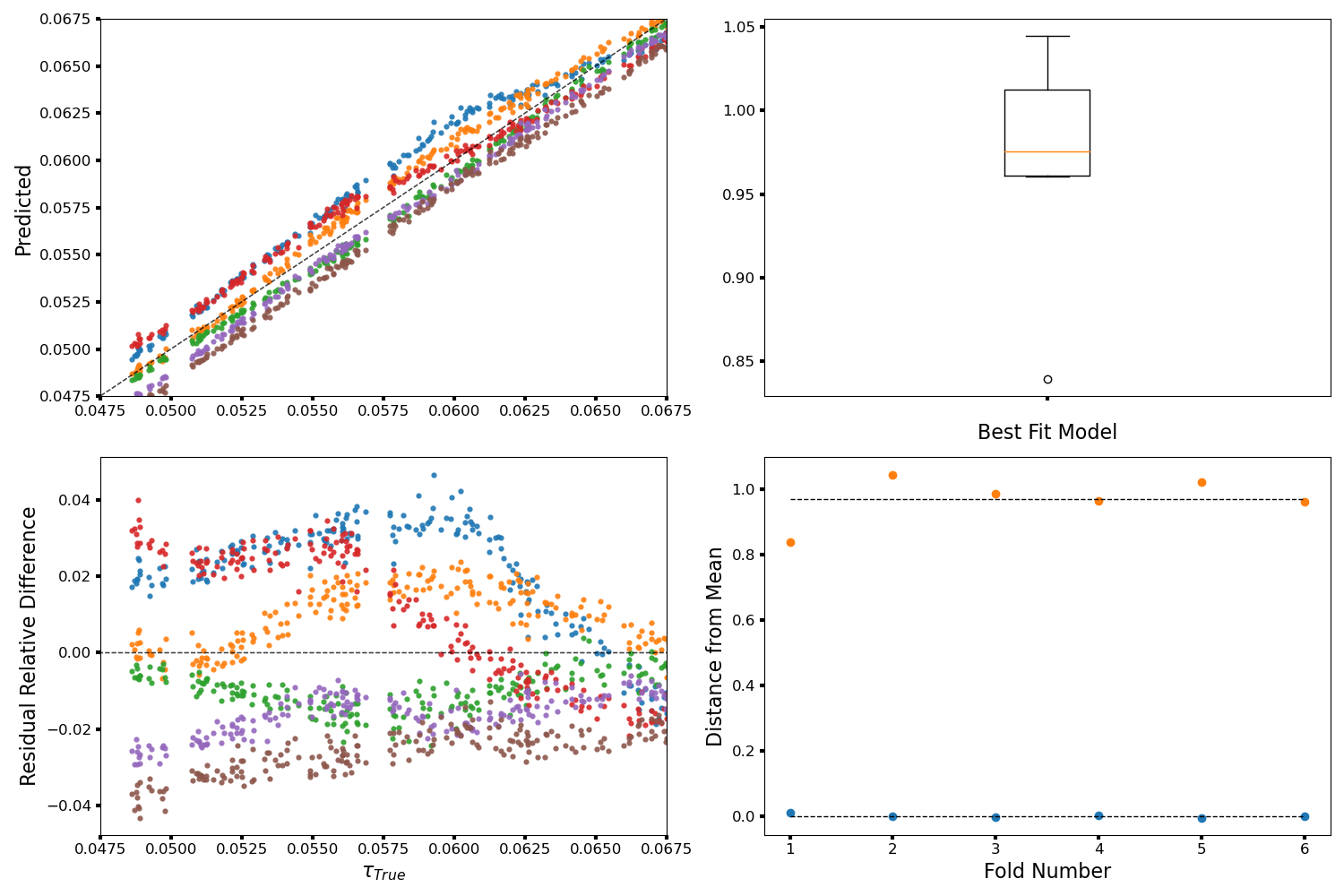}
  \caption{A visualization of the performance of the best-performing CNN on the ``Full'' dataset. Top left: a scatter plot of the ``true'' value $\tau_\mathrm{true}$ versus the value predicted $\tau_\mathrm{predicted}$ by the trained CNN on the validation data. The different colors and symbols correspond to the 10 different folds we used in our $k$-fold cross-validation (explained more in Sec.~\ref{sec:kfold}). Bottom left: the residual relative difference, defined as $(\tau_\mathrm{predicted} - \tau_\mathrm{true})/\tau_\mathrm{true}$, which when squared is used as the loss function for training. Top right: the slope and intercept of a linear fit to the performance of a trained model. For an unbiased network, the intercept has a value of 0 and the slope has a value of 1 and the standard deviation for this value is 0.0267. Bottom right: a box plot of the slope (orange) of the trained network across the different folds. The average value, 0.9694, is nearly 1.0, though there is a significant low outlier whose slope is significantly less than 1 (the single point below the box). See Sec.~\ref{sec:results} for further discussion.}
  \label{fig:ResidualPlots_nowedge}
\end{figure*}

Figure~\ref{fig:ModelComplexity} shows model performance as a function of model complexity. The $x$-axis shows the number of trainable parameters, and the $y$-axis shows the variance of the loss function after performing 10-fold cross-validation. 
Networks with low complexity showed relatively high variance in their average loss function values, most likely indicating that they lacked sufficient flexibility to model the data accurately. Above a certain threshold of about 400,000 trainable parameters, the variance does not decrease significantly. 
This could indicate that there are insufficient
training data to adequately make use of the increase in model complexity, or that
the additional number of parameters is not be necessary to accurately capture the behavior of the input data.

Table~\ref{table:ModelSummary} details the final architectures of our networks arrived at by this hyperparameter optimization. We chose the model that showed the smallest variance as the ``best'' for the purposes of evaluating. We did this for both the ``Full'' and ``Cut'' datasets, which yielded slightly different network architectures. From left to right, the columns show the architectures of the model in \cite{laplante_ntampaka2019}, the model trained on the complete data, and the model that was trained on data where foreground-contaminated $k$-modes were removed from the data. Interestingly, both the ``Full'' and ``Cut'' networks are more complex than the model used in \cite{laplante_ntampaka2019}, but are slightly different from each other.

\subsection{$k$-fold Cross Validation}
\label{sec:kfold}
\begin{figure*}
  \centering
  \includegraphics[width=1.0\textwidth]{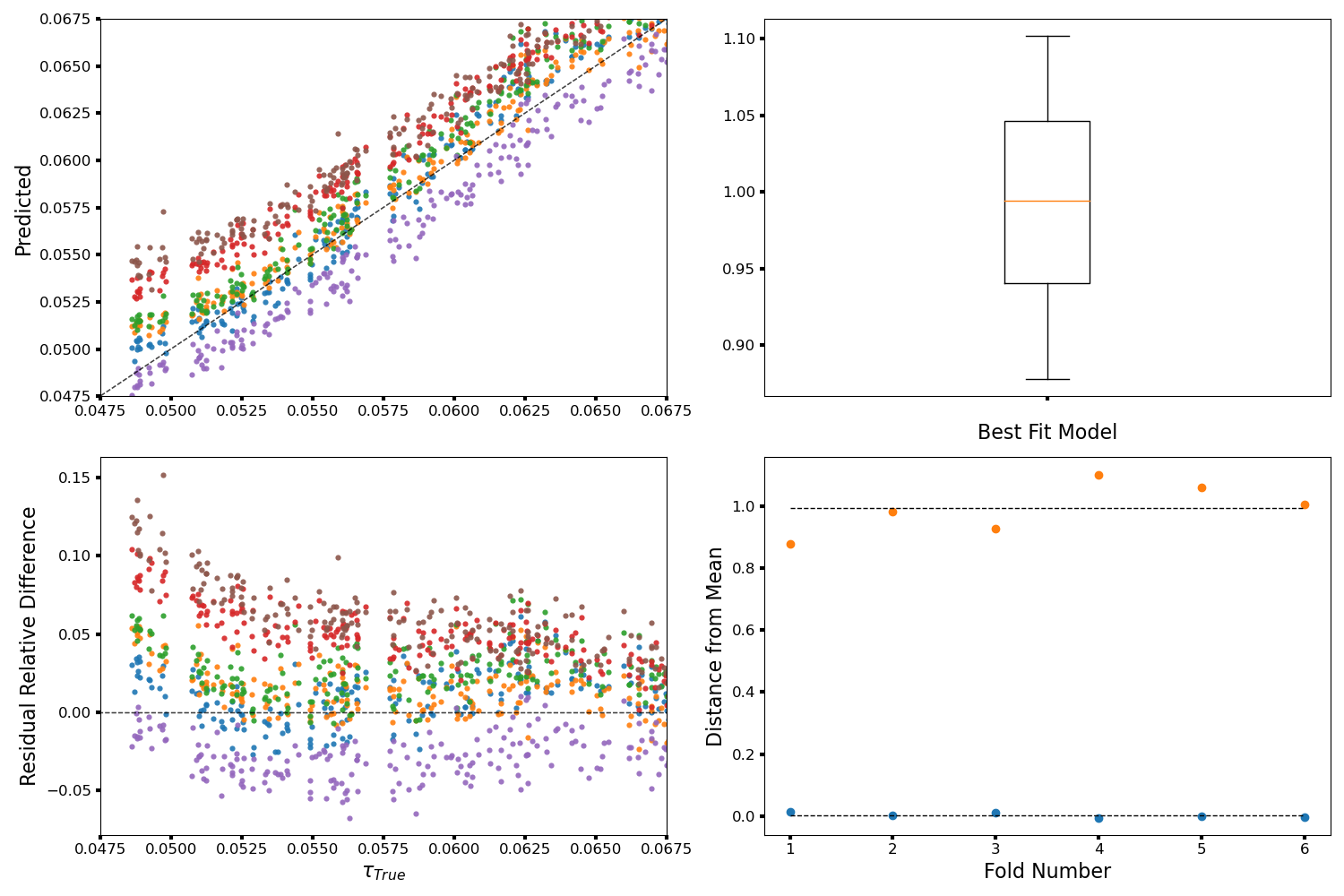}
  \caption{A visualization of the performance of the best-performing CNN on the ``Cut'' dataset. The panels are the same as in Figure~\ref{fig:ResidualPlots_nowedge}. Note that for these models, the variance in the predictions is higher, and there is a slight bias in the slope where the average value is 0.9926 and the standard deviation for the slope value is 0.0308. See Sec.~\ref{sec:results} for further discussion.}
  \label{fig:ResidualPlots_wedge}
\end{figure*}

Cross-validation is another way of ensuring robustness in the model at the expense of computation. In order to train a CNN model, one popular technique is to split the data into training and testing. The ratio of the split can be 90\% training data and 10\% testing data (referred to as 90/10), or 80/20. 
In this work, we use $k$-fold cross validation to help demonstrate that the performance of the trained CNN model does not vary significantly between different partitioning of the input data. To accomplish this, we reserve a randomly selected 20\% of the total pool of images as ``test'' data that is not used for training or validation. Then, we use six-fold validation within the training set. We divide the data up into six equally sized groups, and train ten different networks. Each network uses a different group in turn to serve as validation data in the training process. Throughout the work, results we show are for predictions made on the ``test'' data that was not used as training or validation data. We also use results from different folds as estimates of the bias term in Equation~(\ref{eq:finalgeneralizationerror}). For each fold, we perform a linear regression of the predicted versus true values of $\tau$. We then compute the slope and $y$-intercept of these lines. These can be interpreted as multiplicative and additive forms of bias, and for a well-trained model, the values should be 1 and 0, respectively. Deviations from these values can indicate that a particular CNN model is not well-suited for the problem at hand resulting in a model that does not generalize to data not present in the training set. To remedy this, the hyperparameters may require adjustment as described above in Sec.~\ref{sec:hyperparameters}. Estimating the variance and noise terms is also important, though not at all quantified by fitting the slope and intercept of a linear regression model. We discuss means by which these forms of error can be quantified below in Sec.~\ref{sec:results}.

\begin{figure}
  \centering
  \includegraphics[width=1.0\textwidth]{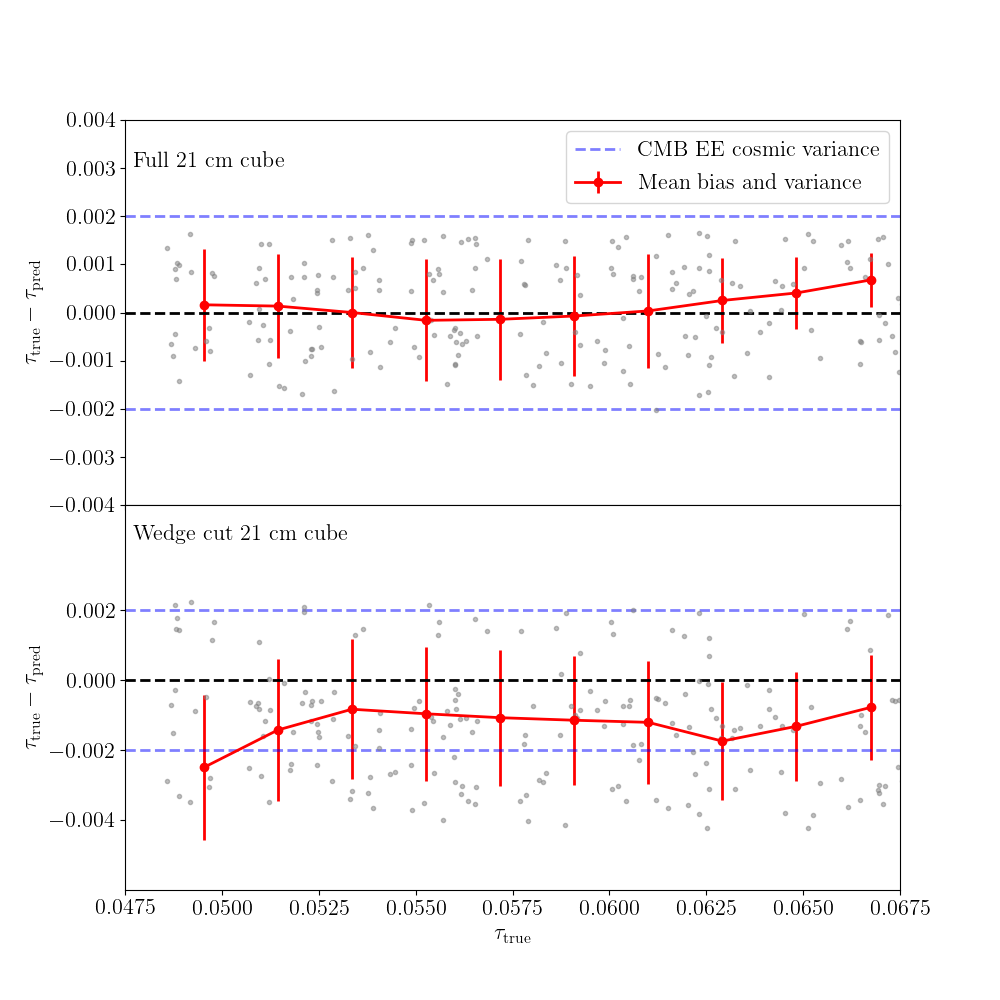}
  \caption{The accuracy with which our machine learning-based approach is able to determine the value of $\tau$. Note the values on the $y$-axis are absolute differences, rather than relative ones as in Figures~\ref{fig:ResidualPlots_nowedge} and \ref{fig:ResidualPlots_wedge}. Also shown are uncertainties associated with sample-variance-limited measurements of $C_\ell^{EE}$ \protect\cite{reichardt2016}. As can be seen, our method produces results that are typically better than what can be obtained from the CMB alone, even for the case where foreground-contaminated $k$-modes have been removed from the dataset. Note that the error bars shown are empirically derived from the training data, and are not ``proper'' error bars in either the Bayesian or Frequentist sense. See Sec.~\ref{sec:tau_comparison} for additional discussion.}
  \label{fig:STD_of_ResidualPlots}
\end{figure}


\begin{figure}
  \centering
  \includegraphics[width=1.0\textwidth]{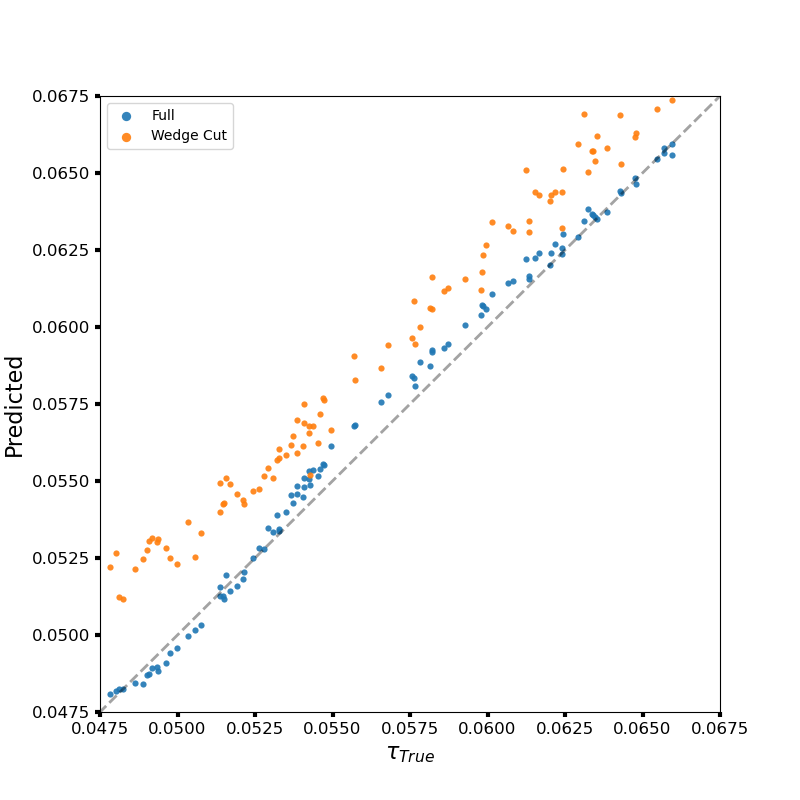}
  
  \caption{Extrapolation of the current models to make predictions on simulated data with a different cosmology generated using the 9-year results from \textit{WMAP}. We are still able to extract $\tau$ for different cosmologies.}
  \label{fig:wmap9_predictions}
\end{figure}

\begin{figure}
  \includegraphics[width=\textwidth]{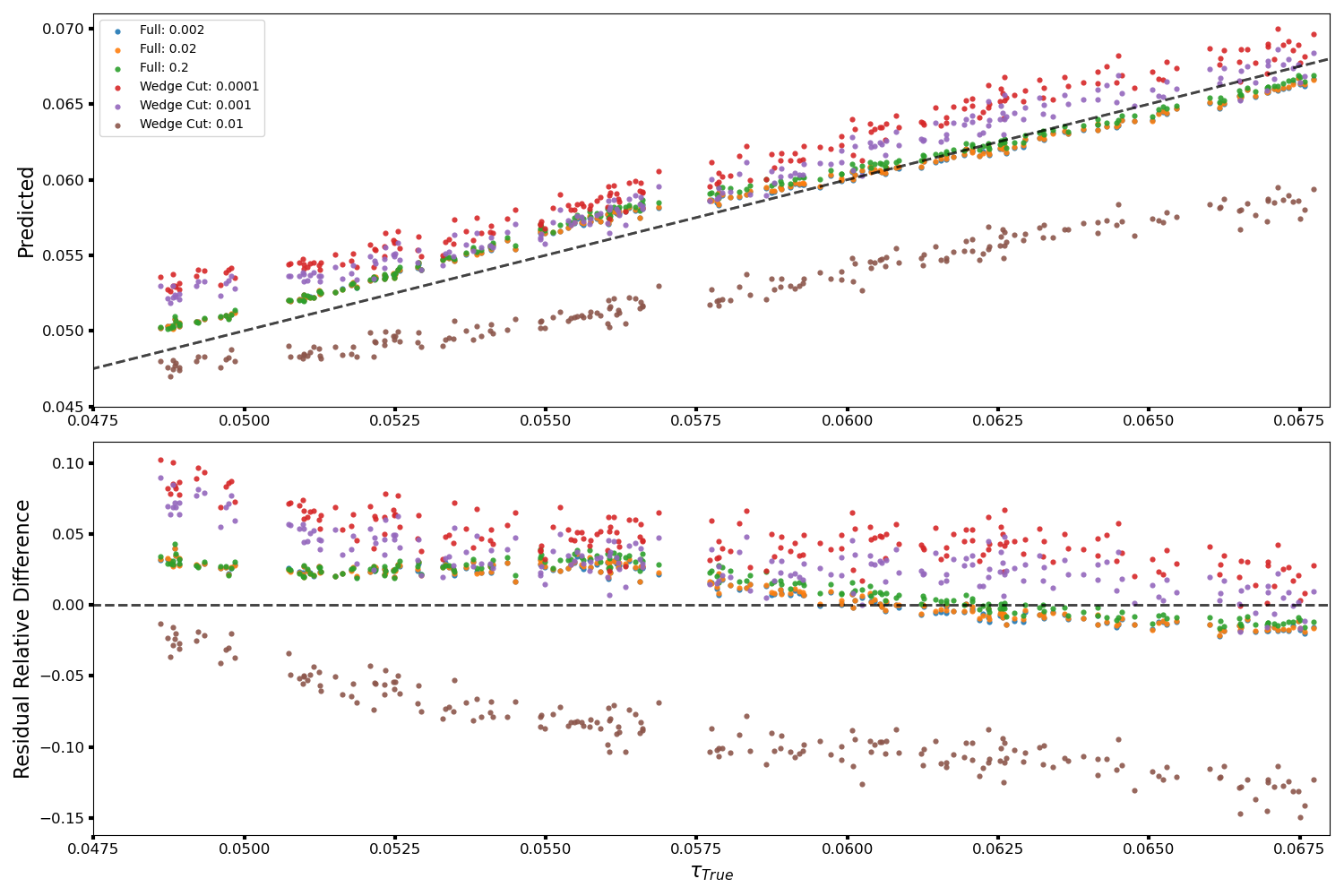}
    \caption{Noise investigation of simulated data generated from Planck 2018 cosmologies. The analysis is conducted by adding white Gaussian noise to input test image data with mean zero and variance set at 0.1, 0.01, and 0.001 of the typical variance of the images. 
    The top panel shows the prediction performance of the CNNs trained either on full data or wedge cut data with no noise when predicting on noisy data. Models that are unbiased will make predictions that follow the one-to-one black dashed line. The bottom panel shows the relative model residuals.}
  \label{fig:noise_planck}
\end{figure}

\subsection{Results of Regression}
\label{sec:results}
 

Figures~\ref{fig:ResidualPlots_nowedge} and \ref{fig:ResidualPlots_wedge} describe the performance of the CNN regression for $\tau$ for both models trained on data without the foreground-contaminated and the model with the $k$-modes removed. These neural networks provide an estimate of $\tau$. The top left corner plot details the one-to-one relationship between the true and predicted tau values. The various colors and symbols represent the 10 different folds used in our $k$-fold cross-validation (discussed more above in Sec.~\ref{sec:kfold}). The bottom left describes the relative difference between the true and predicted value. Note that when squared, this quantity is used as the loss function, written explicitly in Equation~(\ref{eq:RSR}). The top right and bottom right plots describe 10 different slopes of the one-to-one lines observed in the top left plot. As discussed above in Secs.~\ref{sec:biasvar} and \ref{sec:kfold}, these quantities provide an estimate of the bias from the bias-variance tradeoff. For an unbiased CNN, the slope has a value of 1 and the intercept has a value of 0. As can be seen in the different figures, the relationship between the predicted values of $\tau$ and the true values of $\tau$ are quite linear. More specifically, the one-to-one relation between the predicted values and true values show strong positive correlation. With both fully trained CNNs, we were able to recover values to better than $<3.06\%$ percent precision.

\begin{figure*}
  \centering
  \mbox{}
  \includegraphics[width=0.49\textwidth]{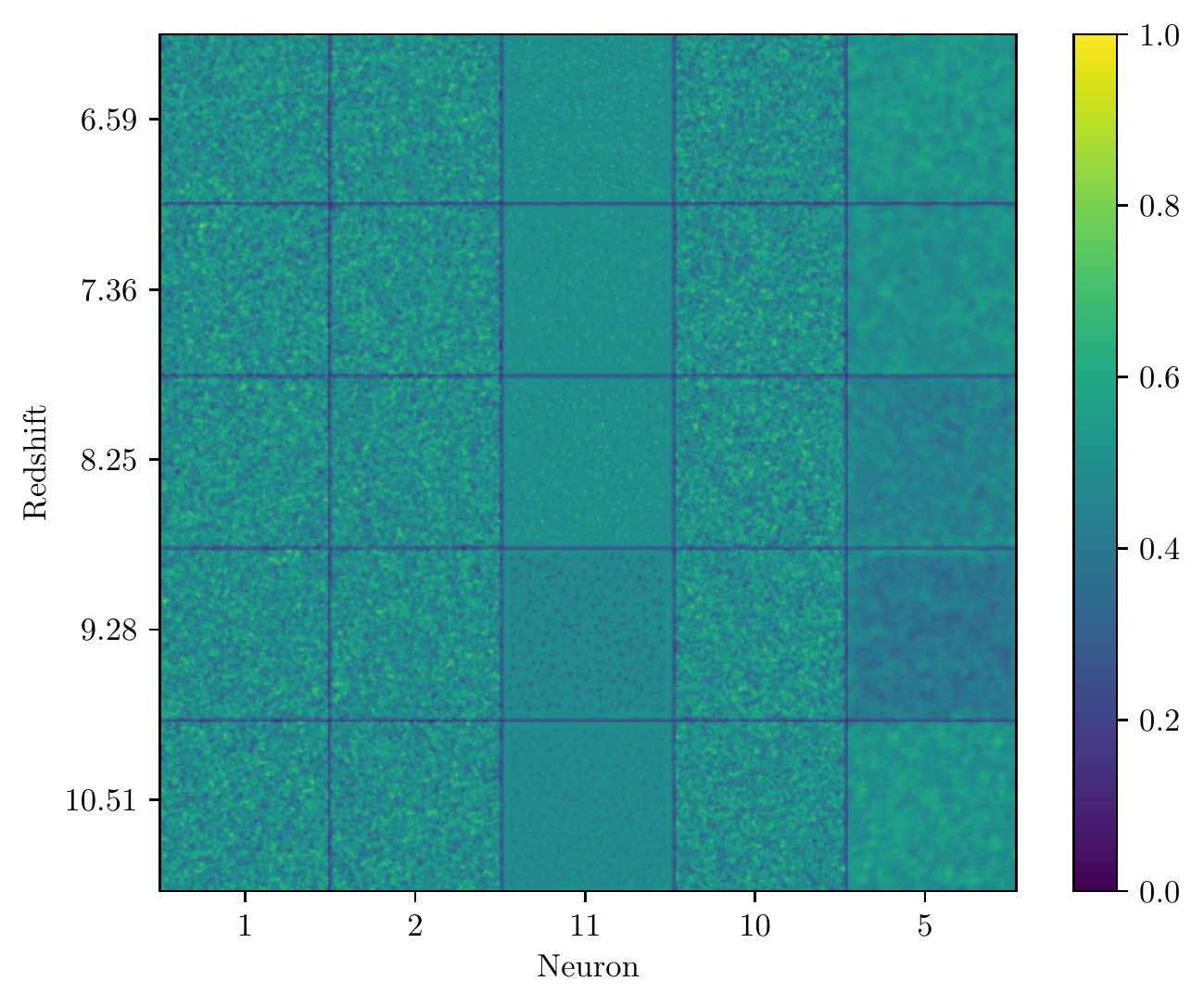}\hfill
  \includegraphics[width=0.49\textwidth]{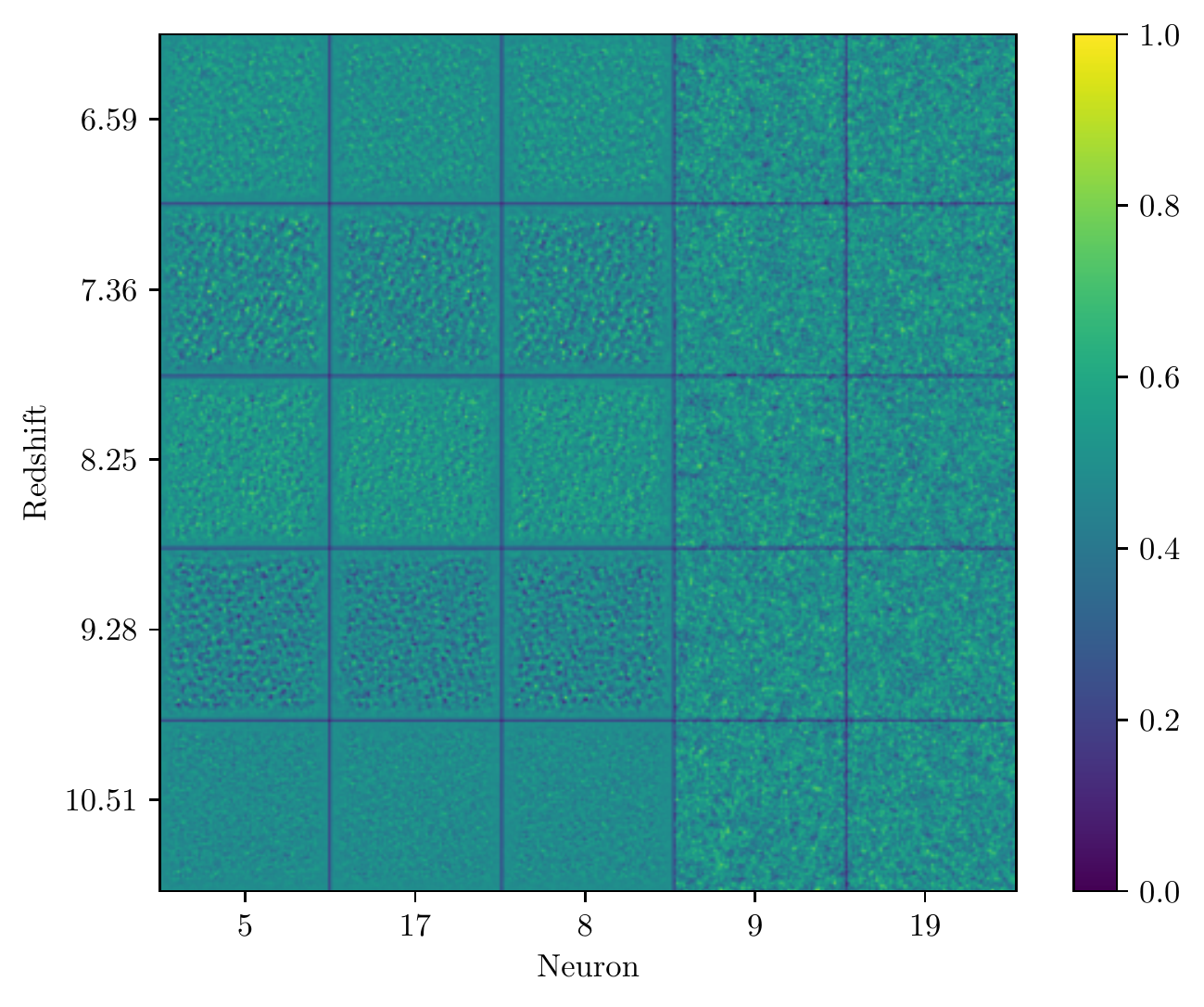}
  \mbox{}
  \caption{A visualization of the activation maximization technique for the 5 neurons most strongly connected to the final output neuron for the ``Full'' data (left) and the ``Cut'' data (right) networks from the best performing model fold. The neurons are organized by column and rank-ordered from left-to-right by the magnitude of the weight connecting them to the final neuron. The different rows represent redshift values corresponding to the different slices of the input data. See Sec.~\ref{sec:visualization} for additional discussion.}
  \label{fig:layers_act_max}
\end{figure*}

The scatter plots in the top- and bottom-left panels show the full relationship between the true value and predicted value of $\tau$. In both the ``Full'' and ``Cut'' networks, there are slight systematic biases where small values of true $\tau$ are biased low, intermediate values of $\tau$ are biased high, and the highest values of $\tau$ are biased slightly low. However, as can be seen in the plots, the bias is typically smaller than the scatter in the values, and so the true values of $\tau$ are generally included as part of the scatter. The top- and bottom-right plots of the figures emphasize the bias component of the trained network: the value of the slope is a proxy for the multiplicative bias inherent in the network, and the intercept is an additive bias. In contrast to Figure~\ref{fig:bias_var_ex}, the bias of the lines is generally small, even across different folds. These results suggest that the networks described in Table~\ref{table:ModelSummary} are sufficiently complex to capture the important features of the input data, and are correctly minimizing the variance and noise terms while providing an unbiased estimate of $\tau$.

As a way of determining the bias and variance of the predicted values as a function of the input value of $\tau$, we combine the predicted values of $\tau$ across the 10 folds of our input data. We then divide the true values of $\tau$ in the input dataset into ten discrete bins of equal width. Within each bin, we compute the mean value and the standard deviation. The mean value is a proxy for the bias (because a mean value different from the true value denotes a biased estimator) and the standard deviation encapsulates the variance in the output of the model predictions as well as the noise in the trained model. In general, we find that the variance tends to be several factors larger than the bias. While this result is not as ideal as having a truly unbiased estimator, it does mean that the variance by itself is a reasonable approximation to the total error of the trained network.

Figure~\ref{fig:STD_of_ResidualPlots} shows the results of performing such an analysis for both the ``Full'' and ``Cut'' datasets. As noted above, the results are slightly biased as a function of $\tau$, and the direction of the bias changes as a function of the input value. However, as can also be seen, this bias tends to be smaller than the variance in the output values, and so the mean predicted value of $\tau$ tends to be consistent with the proper value within 1$\sigma$ of the empirical standard deviation. 
As a point of comparison in Figure~\ref{fig:STD_of_ResidualPlots}, we show the best-case error estimates that can be provided on $\tau$ from measurements of the CMB alone. We further discuss this comparison below in Sec.~\ref{sec:tau_comparison}.

\section{Discussion}
\label{sec:discussion}


        


\subsection{Visualizing CNN Feature Extraction}
\label{sec:visualization}

When using image-based machine learning techniques such as CNNs, an interesting question is how to interpret the inner workings of the algorithm. 
One way to do this is to examine the effect on an input image of the different convolutional filters at each layer.  Though the resulting ``images'' no longer represent information in the same space as the input images after the first input layer,
they do contain information about which particular features of the map the CNN has learned to focus on. Alternatively, one can use the \textbf{activation maximization} technique \cite{erhan_etal2009} to visualize the important features in the input map directly, rather than using a partially processed image. In this approach, a specific neuron in a dense layer or convolution filter in a convolutional layer of a trained network is chosen. An initially random input image is 
gradually transformed into an image that maximizes the response of the chosen neuron or filter layer through gradient ascent. The resulting input images do not necessarily look like input images, but instead emphasize the features that are important for the machine to discriminate between different values or feature classes. Such an approach helps visualize which aspects of the the image are being used by the trained network to provide predictions, and complement other methods of visualizing CNN operations. To carry out the actual computation, we make use of the \texttt{keras-vis}\footnote{\url{https://github.com/raghakot/keras-vis/}} package.

We applied the activation maximization technique to the fully trained networks developed in this work. This approach is sometimes employed for a classification problem, where the resulting images can be interpreted as the features that are most important for categorizing an imagine into a particular class. However, for a regression problem like the one at hand, these instead show features that lead to a large response of a particular neuron, typically one deep in the network. Though not as clear an indicator of the network's response as in a classification problem, the resulting images nevertheless contain features that the network has identified as ways to distinguish different output values. In both the ``Full Modes'' and ``Cut Modes'' architectures described in Sec.~\ref{sec:cnn}, there is a 20-neuron dense layer immediately before the final prediction neuron for the value of $\tau$. After training the network, we examine the magnitude of the weights connecting these 20 neurons and the output neuron. In general, the larger the magnitude of the weight connecting these neurons (positive or negative), the more influence the individual neuron has on the output prediction. For the two different networks, we identified the five neurons that had the largest magnitude connection to the output neuron. We then used the activation maximization technique to generate input images which would maximize the response of these neurons. The resulting images have the same dimensionality as the input images, and in particular have 30 ``color channels'' which correspond to the redshift layers of the input data.

Figure~\ref{fig:layers_act_max} shows the five most strongly connected neurons for various input redshift values for the ``Full Modes'' and ``Cut Modes'' networks, respectively. The columns show the maximal input for different neurons, rank-ordered from left-to-right by the magnitude of the weight connecting them with the final output neuron. The different rows correspond to the same redshift layers in the input data. When comparing the features between the different networks, several different trends emerge. First, for an individual neuron, the features that appear in the input images are similar for different redshifts.
Because different input images are comparable between the different input redshifts, this similarity suggests that having many different filter layers initially is important. Multiple filter layers provide sufficient flexibility for identifying various features in the input maps, which are later condensed into features identified by hidden layers deeper into the network. Also of interest is the fact that generally, the features seem to be contrasts of large and small values at different scales, which roughly correspond to the size of individual ionized regions when viewing unprocessed input images. This result suggests that the CNN may be using the size of ionization bubbles at different redshifts to inform the overall value of $\tau$, though we caution that such a one-to-one mapping is not necessarily faithful to the actual operations being performed by the CNN.

When comparing the features identified in the ``Full Modes'' versus ``Cut Modes'' networks, there are several interesting differences. Of particular note is the features that the two different architectures treat as the ``most important'' in terms of informing the overall output value. The most important maps for the ``Full'' are qualitatively similar to the ``Cut'' network, but they are not the most important. Instead, the ``Cut'' network seems to be identifying features that are deviations from a background level (typically either higher, seen in the bright yellow regions, or lower, seen in the dark blue regions) rather than high-low variations near each other. Accordingly, these features appear non-Gaussian, perhaps emphasizing that the 21\,cm maps are highly non-Gaussian (especially so with the large-scale contaminated modes removed). As such, the CNN appears to be making use of important information that is difficult to capture in the form of summary statistics, which bolsters the claim that CNNs can complement more traditional methods of analyzing image-based data.

\subsection{Comparison with Limits on $\tau$ from Other Methods}
\label{sec:tau_comparison}

There are a number of well-established techniques for measuring $\tau$.
The current best constraints come from using CMB data, such as the all-sky temperature auto-power spectrum (denoted $C_\ell^{TT}$), as well as the large-angle auto-power spectrum of gradient-like E-modes (denoted $C_\ell^{EE}$). $C_\ell^{TT}$ is sensitive to the combination of parameters $A_S e^{-2\tau}$, where $A_S$ is the initial amplitude of scalar perturbations. This degeneracy can be partially broken by using CMB lensing maps. Alternatively, the low-$\ell$ portion of $C_\ell^{EE}$ follows a rough scaling of $C_{2\leq\ell\leq 20}^{EE} \propto \tau^2$ \cite{page_etal2007}, which provides an additional means of determining $\tau$. The Planck 2015 set of cosmological parameters \cite{planck2015} reports a value of $\tau = 0.066 \pm 0.016$, or a roughly 25\% uncertainty. The Planck 2018 results \cite{planck2018} find a value of $\tau = 0.054 \pm 0.007$, about a 13\% uncertainty.  Other experiments, such as the EDGES high-frequency instrument, have further been able to place upper limits on the value of $\tau$ consistent with the measurements of Planck \cite{monsalve_etal2019}. In principle, measurements of $C_\ell^{EE}$ can provide much tighter constraints on $\tau$ than $C_\ell^{TT}$. However, due to sample variance, these measurements cannot provide an uncertainty better than $\sigma_\tau \sim 0.002$ \cite{reichardt2016}, which corresponds to a roughly 4\% uncertainty. These measurements are projected to be made with future space-based CMB instruments, such as LiteBIRD \cite{litebird} and Pixie \cite{pixie}, which are not scheduled to fly until well into the next decade.

Figure~\ref{fig:STD_of_ResidualPlots} shows the accuracy with which our machine learning-based approach is able to determine the value of $\tau$, along with the sample variance possible from $C_\ell^{EE}$. As can be seen, the accuracy of our method is typically better than what can be obtained from the CMB alone, even for the case where foreground-contaminated $k$-modes have been removed from the dataset. Some important caveats remain, however. Importantly, the error bars shown in Figure~\ref{fig:STD_of_ResidualPlots} are empirically derived from the training data, and are not ``proper'' error bars in either the Bayesian or Frequentist sense. Nevertheless, the error bars are an indication that the value of $\tau$ inferred from this method is smaller than what is possible from the CMB alone, and is a promising tool to use in conjunction with more traditional methods. At the same time, further work is required to understand the impact the training data has on correctly inferring the value of $\tau$, either due to the quantity of training data or the semi-analytic model used to generate it. We plan to investigate these effects in future studies.

The results here are, of course, preliminary and should not be treated as a proper forecast of the potential accuracy of future 21\,cm experiments.  While we have included the effect of lost modes due to foreground contamination, we have not included the effect of other systematic errors in the 21\,cm measurement on the result. In addition, the analysis here does not consider realistic instrument noise which varies with respect to the cosmological $k$-mode, which will naturally increase the error bars \cite{pober_etal2014}. Working against this, our network only works on a single FoV of $\sim 10^{\circ}$, whereas HERA will sample approximately 10 such non-overlapping fields over some 1000 square degrees. We may also be able to use a fewer number of frequency channels to obtain comparable results, which will allow for generating multiple spectral windows to improve sensitivity. A forecast for more realistic systematic and sensitivity calculations will be presented in future work.

Another point of comparison for the ability to infer the value of $\tau$ is the analysis in \cite{liu16}. The approach taken in that paper was to treat $\tau$ as a parameter to be inferred jointly with other CMB parameters, such as $\Omega_c$ and $\sigma_8$. In that case, the final marginalization over $\tau$ and other parameters yielded an uncertainly of $\sigma_\tau = 0.0016$, or about 3\%. This is comparable to the uncertainty for our ``Cut'' model, and larger by roughly a factor of 50\% compared to our ``Full'' model, as seen in Figure~\ref{fig:STD_of_ResidualPlots}. Note that in our approach, the background cosmology was assumed to be fixed, and we do not attempt to jointly constrain the value of $\tau$ in concert with the other cosmological parameters. Performing a joint fit for other cosmological parameters is computationally intensive, and requires the use of cosmological emulators \cite{Kern_2017} or other techniques to accelerate the forward-modeling component. In future work, we plan to use Bayesian neural networks (BNNs) to provide more robust distributions of the errors associated with machine learning modeling. Future directions may also include varying the background cosmology to understand the uncertainty associated with inferring $\tau$ using the 21\,cm alone and how sensitive these measurements are to other parameters changing.

\subsection{Testing the Effects of Different Cosmology and Noise}
\label{sec:noise_cosmo}

The networks above were trained on 21 cm data generated using Planck 2018 (Planck18) cosmology. From previous work showing only a weak dependence of the 21 cm power spectrum on cosmology (e.g., \cite{Kern_2017}) we can similarly expect that the dependence of $\tau$ on cosmological parameters is weak.  
To provide an estimate of the kinds of errors which would occur in this analysis if the underlying cosmology is wrong, we generated a new test data set using Wilkinson Microwave Anisotropy Probe (WMAP) 9 year results \cite{hinshaw_etal2013}.  The most notable difference between these cosmologies ($\tau$ aside) is $\Omega_m$, which differs by $\sim$10\%.   We then used the networks trained on Planck18 to make predictions on the WMAP-9 data. In Figure~\ref{fig:wmap9_predictions}, we show the results.
For the case of the full data, the predictions are nearly as good as using the correct cosmology.
Interestingly, the model trained on wedge cut data makes optical depth predictions that are biased high in this new cosmology, though the bias is only slightly larger in magnitude than was observed in Figure \ref{fig:STD_of_ResidualPlots}.   Given that the tight correlation remains, it seems reasonable that a fuller analysis which properly marginalized over the cosmological parameter uncertainty would not increase the prediction errors unduly.

While the actual noise of 21 cm instruments will be quite complicated, we can gain some insight into the robustness of this method to noise by simply adding mean zero white Gaussian noise to the test image data and re-running the predictions. 
We chose the variance to be 0.1, 0.01, and 0.001 of the typical variance of the Planck18 cosmology images.
Figure~\ref{fig:noise_planck} shows the ability of the network to predict the optical depth at these different noise levels. 
The predictions follow the one-to-one line closely, except for the highest noise level of wedge-cut data, which shows a noticeable bias.  However, even in this case, the clear correlation between $\tau_{true}$ and $\tau_{pred}$ remains, giving confidence that a network properly trained using the actual noise properties of the instrument would still be able to make accurate predictions. 

\section{Conclusion}
\label{sec:conclusion}
In this paper, we show that we are able to train two different Convolution Neural Networks on simulated data, one with the all Fourier $k$-modes included and the other without foreground-contaminated $k$-modes removed from the data. Through the use of hyperparameter optimization, we are able to find model architectures that are best suited for each application, and perform well over the full range of input data. We demonstrated that we can make accurate $\tau$ predictions using networks trained on both simulations types reasonably well. These simulated input images reflect the effects of the foreground avoidance strategy implemented by HERA as part of data processing. If some of these foreground modes can be used instead of discarded, the ultimate performance may be closer to the full data set than the one with the $k$-modes removed. We show that we are able to provide constraints on $\tau$ with a fractional error of $3.06\%$ or better, which makes this approach competitive with low-$\ell$ observations of the CMB auto-power spectrum $C_\ell^{EE}$. Due to the fact that instruments capable of providing such a constraint are many years away, using 21\,cm measurements may be able to provide a constraint on a shorter time line.

Machine learning techniques such as that outline here are most powerful in conjunction with more traditional analyses, 
providing additional cross-checks of results inferred by other means.
In future work, we plan to make use of Bayesian neural networks (BNNs) to provide robust error estimates in addition to the predicted values for a particular CNN model. These novel analysis methods can supplement other established methods, and help bolster confidence in inferences made through other analysis techniques.

\section*{Acknowledgments}
We thank Adrian Liu for helpful discussions about this work. T.S.B. and J.E.A. acknowledge support from NSF CAREER award AST-1455151. This material is based upon work supported by the National Science Foundation under Grant No. 1636646, the Gordon and Betty Moore Foundation, and institutional support from the HERA collaboration partners. HERA is hosted by the South African Radio Astronomy Observatory, which is a facility of the National Research Foundation, an agency of the Department of Science and Technology. This work used the Extreme Science and Engineering Discovery Environment (XSEDE), which is supported by National Science Foundation grant No. ACI-1548562 \cite{xsede2014}. Specifically, it used the Bridges system, which is supported by NSF award No. ACI-1445606, at the Pittsburgh Supercomputing Center \cite{bridges2015}.


\section*{References}

\bibliographystyle{jphysicsB}
\bibliography{references}

\end{document}